\documentclass[aps,prb,twocolumn,showpacs]{revtex4-1} 

\usepackage{graphicx}
\DeclareGraphicsExtensions{.png,.jpg,.eps,.pdf}

\usepackage{xcolor}
\usepackage{hyperref}
\usepackage{amsmath}
\usepackage{amssymb}
\usepackage{bm}

\usepackage{amsthm}
\usepackage{bm}
\usepackage{amsfonts}
\usepackage{soul}

\begin{document}

\title{Electrically-detected single-spin resonance  with Quantum Spin Hall edge states}

\author{F. Delgado$^1$ and J. Fern\'andez-Rossier$^{2,3}$}
\address{$^1$ Instituto de estudios avanzados IUDEA, Departamento de F\'{i}sica, Universidad de La Laguna, C/Astrof\'{i}sico Francisco S\'anchez, s/n. 38203, La Laguna}

\address{$^2$ International Iberian Nanotechnology Laboratory (INL),
Av. Mestre Jos\'e Veiga, 4715-310 Braga, Portugal
\footnote{On permanent leave from Departamento de F\'{i}sica Aplicada, Universidad de Alicante, 03690 San Vicente del Raspeig, Spain}$^,$\footnote{joaquin.fernandez-rossier@inl.int}
}

\date{\today}

\begin{abstract}
Detection is most often the main impediment to reduce the number of spins in 
paramagnetic resonance experiments. Here we propose a new route to carry out electrically-detected spin resonance of an individual spin, placed at the edge 
 of a quantum spin Hall insulator (QSHI).  The edges of a QSHI host a one dimensional electron gas with perfect  spin-momentum locking. 
Therefore, the spin relaxation induced by emission of an electron-hole pair at the edge state of the QSHI can generate 
 current. Here we demonstrate that driving the system with an $AC$ signal, a nonequilibrium occupation  can be induced in the absence of applied bias voltage, resulting in a $DC$ measurable current. 
We compute the $DC$ current as a function of the Rabi frequency $\Omega$, the spin relaxation and decoherence times, $T_1$ and we discuss the feasibility of this experiment with state of the art instrumentation.
\end{abstract}

\maketitle


\section{Introduction}
%
%
The sensitivity limit of   commonly available electron paramagnetic resonance (EPR)
spectrometers is in the range of $10^{13}$ spins \cite{brustolon09}.
This number can be dramatically reduced in taylored set-ups \cite{Probst_Bienfait_apl_2017}.
In some special systems, such as NV-centers, permit one to carry out single-spin resonance using optical readout, made possible both by the fact that NV centers are very good single-photon emitters and their photon yield is spin dependent.~\cite{gruber97}  Using spin-to-charge conversion,  electrically detected 
single spin resonance has been demonstrated for defects in field effect-transistors,~\cite{xiao04} quantum dots\cite{koppens06,pioro08} and single dopants in silicon.~\cite{pla12}
 Electrically detected single spin resonance  with subatomic spatial resolution has been also demonstrated\cite{Baumann_Paul_science_2015} using scanning tunneling microscopy (ESR-STM). 

\begin{figure}[t!]
 \centering
 \includegraphics[width=1.\linewidth]{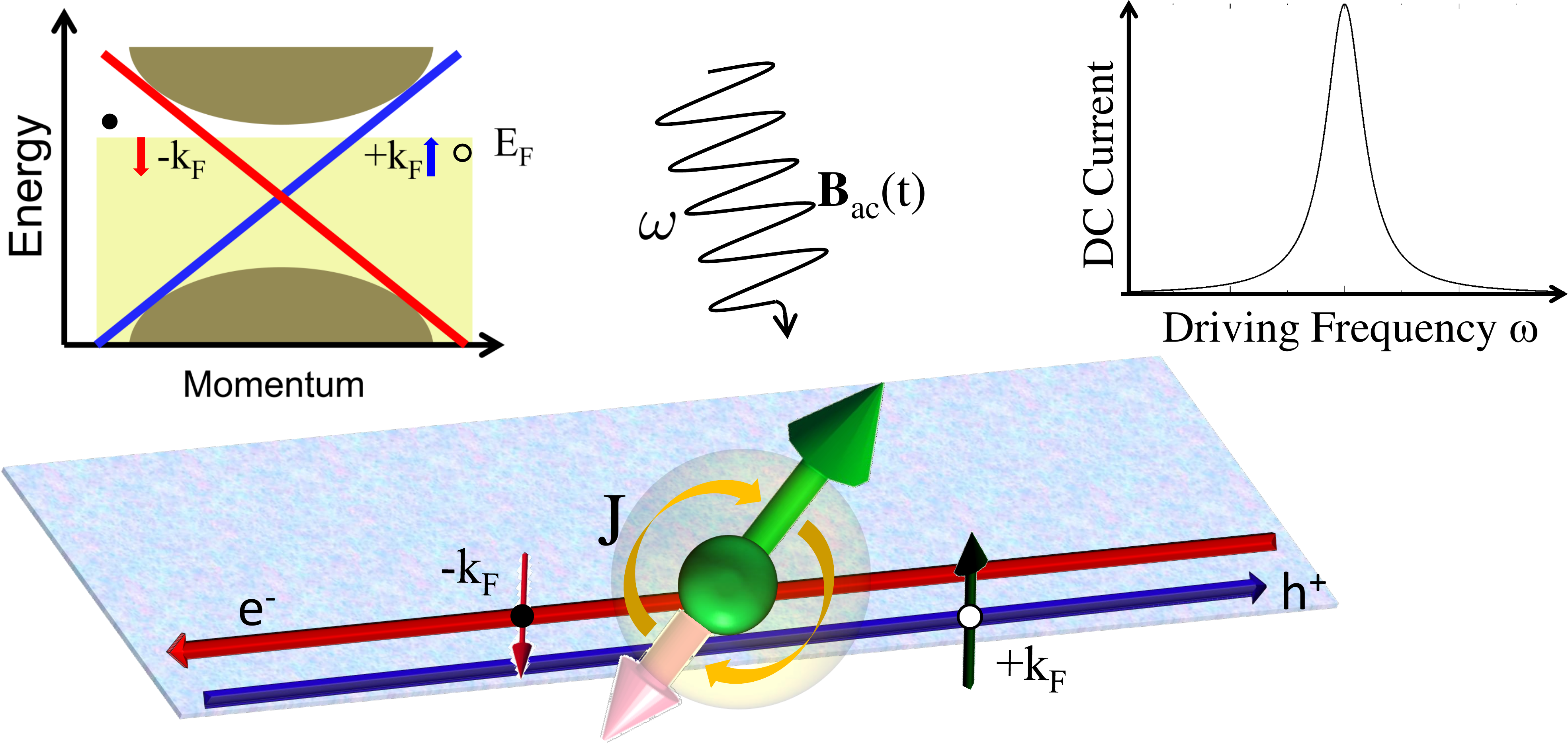}
\caption{Scheme of the device proposed for the electrically-detected single spin detection. A local spin $S$ is exchange coupled to one edge state of a QSHI, where momentum direction and spin orientation are locked, see left inset. When the system is under the action of an external $AC$ magnetic field with frequency $\omega$ and intensity  determined by the Rabi rate $\Omega$, a frequency-dependent non-equilibrium steady state occupation is established (see right inset), where one of the spin transitions $\Delta S_z$ is favored. This, in turns, leads to a net $DC$ electrical current along the QSHI edge.  
}
\label{scheme}
\label{fig1}
\end{figure}

 Here we explore  the spin-locked edges states of a two dimensional Quantum Spin Hall insulator\cite{Kane_Mele_prl_2005,wu06}  (QSHI) to accomplish the electrical readout of the spin  resonance of an individual spin sitting on the edge.  The edge states of QSHI are predicted to have 
a 
 one-to-one relation between the propagation direction and the spin orientation along a system-dependent spin quantization axis, see Fig.~\ref{scheme}. As a result, pumping spin along this axis
  entails electrical current flow.  As we discuss below, if a externally pumped  localized spin is exchange coupled to the spin-locked edges,  it will generate a $DC$ current.
 
 Experimental evidences of the spin-locked edge states in QSHI are indirect.  In the absence of magnetic impurities,  edge states should have no backscattering and therefore a quantized conductance is expected.~\cite{Bernevig_Zhang_prl_2006,Bernevig_Hughes_science_2006,Roth_Brune_science_2009} Values of conductance close to $2e^2/h$ were reported in HgTe/CdTe quantum wells~\cite{konig07} and 
1T' WTe$_2$.~\cite{Peng_Yuan_natcomm_2017,wu18} In addition,
coherent propagation along the edge with scattering properties consistent with strong suppression of backscattering have been observed in bismuth 
bilayers, \cite{Drozdov_Alexandradinata_natphys_2014,Murani_Kasumov_natcomm_2017} and in bismuth nanocontacts\cite{sabater13}. Very relevant for the ensuing discussion,  experiments where magnetic atoms 
interact with edge states in bismuth bilayer and produced back-scattering have been demonstrated \cite{Jack_Xie_pnas_2020}.

 The interplay between local spins and the spin-locked edge states of a QSHI has been widely studied theoretically~ \cite{maciejko09,tanaka11,Lunde12,eriksson12,Lunde13,hurley13,narayan13,arrachea15,probst15,silvestrov16,
 vayrynen17,locane17,hsu17,kurilovich19,roura18,Delgado_Rossier_njp_2019,zhuang21}. 
Several physical realizations of the local spin have been considered, including a confined electron in a quantum dot~\cite{probst15,roura18}, nanomagnets,~\cite{arrachea15}   magnetic atoms,~\cite{hurley13,narayan13} spin chains~\cite{Delgado_Rossier_njp_2019}, nuclear spins,~\cite{Lunde13,hsu17,zhuang21}  and  magnetic molecules~\cite{locane17}. Early works focused on the Kondo effect,~\cite{maciejko09,eriksson12} and the influence of magnetic impurities on conductance~\cite{tanaka11}.  More recent works have  addressed    spin-pumping of local moments at the edges by the helical-electron spin current.~\cite{hurley13,narayan13,arrachea15,vayrynen17,probst15,locane17,kurilovich19} 
The reverse problem,  pumping $DC$ current by an external $AC$ excitation of nuclear spins has been addressed recently.~\cite{zhuang21} On a similar standpoint, here we assess whether the paramagnetic spin resonance of an individual spin, in the form of individual magnetic atom,  spin chain or magnetic molecule, could be carried out.

The rest of this paper is organized as follows.  
In Sec.~\ref{secII} we introduce the basic principles of the electrically-detected single spin resonance in a QSHI. An estimation of the maximum $DC$ current is provided in Sec.~\ref{secIII}, while the main limiting factors are discussed in Sec.~\ref{secIV}. Finally, a brief summary  and conclusions is given in Sec.~\ref{secV}.

\section{Electrically detected single-spin resonance in QSHI edge states \label{secII}}
This work builds on the following idea: at
the edge states of QSHI, an electron with spin $-\sigma$ and momentum $-k_F$ can be scattered  to a state with momentum $+k_F$ and spin $\sigma$ by exchange interaction with a local spin.   Here  $\sigma$ is defined along a material-dependent axis that, without  loss of generality, we label as $z$. Since the total spin has to be conserved in the process, the spin change of the electron has to be 
compensated by the spin change of the local magnetic moment. Unless otherwise stated, we only consider a local spin whose spin quantization axis is aligned  along $z$, the same  quantization axis of the  quasiparticle states.   Therefore, for a given local-spin transition with a change of spin  $\Delta S_z= \pm 1$, the quasiparticles undergo a $\pm 2k_F$ backward scattering process with $\Delta \sigma= \mp 1$, on account of their helical nature. 

Crucially, 
the electron-hole pairs  carry a net current whose sign depends on the sign of $\Delta \sigma$.  
The extra electron in one branch and the missing electron in the opposite branch contribute to current flow with the {\em same} polarity. As the edge states are expected to have no back-scattering, the electron and the hole will reach the electrodes and contribute to the current. Hence, if we create an stationary nonequilibrium imbalance in the $\Delta S_z$ transitions by driving the spin transitions with a external $AC$ driving field,  a net $DC$ current will be generated.

We now substantiate the argument mathematically.
Let us take for simplicity a local $S=1/2$ spin moment under the influence of a static magnetic field,
 $B_{\rm eff}=\hbar\omega_0/(g\mu_B)$, where $B_{\rm eff}= B_z + B_z^{\rm other}$ is the sum of the external field $B_z$ and other contributions that could arise from the interaction of the local spin and its environment. Thus, the stationary spin Hamiltonian can be written as
${\cal H}_0= \frac{\hbar\omega_0}{2} \hat \tau_z$, being $\hat\tau_z$ the $z$-Pauli matrix, and 
$\hbar\omega_0=\epsilon_1-\epsilon_0\ge 0$.
 We label $P_{0}$ and $P_{1}$ as the probabilities of occupying the ground ($0$) and excited ($1)$  states, respectively. 
The relevant spin-exchange process that gives place to spin flips is governed by the Hamiltonian:~\cite{Delgado_Rossier_njp_2019}
\begin{equation}
H_{\rm sf}^{\rm QSHI}=
\sum_{k,k'}\frac{J}{2L}
 \left(\hat S^{+} 
L^{\dagger}_k R_{k'} + {\rm h.c.}\right),
 \label{VK_perp}
\end{equation}
where $J$ is the exchange coupling constant, 
$L$ is the length of the edge,
 $L_k^\dag \equiv c_{-(k_F+k),\downarrow}^\dag$ and $R_k^\dag\equiv c_{+(k_F+k),\uparrow}^\dag$ are the left (right) moving fermion operators, 
with $c_{k\sigma}^\dag $ the creation operator 
of a fermion in the edge channel with spin $\sigma$ and momentum $k$. 
Here $\hat S^\pm=1/2(\hat S^x\pm i\hat S^y)$ are the spin-ladder operators.  

 If we define the rate $\Gamma_{1\rightarrow 0}$   and $\Gamma_{0\rightarrow 1}$  of the $\Delta S_z=\mp 1$ 
 process respectively, we can write the electric current flowing to the right as
\begin{equation}
I=2e\left(P_{1}  \Gamma_{1\rightarrow 0} -P_{0}\Gamma_{0\rightarrow 1}\right),
\label{current0}
\end{equation}
where $e$ 
  is the elementary charge and $P_{i}$ are the non-equilibrium occupation of the $i\equiv 0,1$ states.
In equilibrium, the current (\ref{current0}) vanishes because the scattering rates  satisfy the detailed balance principle:
\begin{equation}
\frac{ \Gamma_{1\to 0}}{\Gamma_{0\to 1}}=\frac{P_{0}^{\rm eq}}{P_{1}^{\rm eq}}=  e^{\beta\hbar \omega_0}
\end{equation}
where, $1/\beta=k_B T$ and $P_{i}^{\rm eq}$ are the equilibrium occupations.
We shall now demonstrate that if the local spin is driven away from equilibrium by some external force that does not significantly modify the rates, then a net current can occur. If we write 
$P_i=P_{i}^{\rm eq}\pm\delta P/2$, where the $+$ ($-$) sign corresponds to $i=1$  ($i=0$), 
and taking into account that in thermal equilibrium (without any applied bias voltage) the net current is null, 
then we have:
\begin{equation}
I_{DC}=e\delta P \Gamma_{1\to 0}  \left(1+e^{-\beta \hbar\omega_0}\right).
\label{current1}
\end{equation}
This equation is the starting point of our analysis. It relates the out-of-equilibrium occupations of the two level system and a net current flow. The direction of the current is stablished by the chirality of the spin edge and by the sign of the magnetic field.  For a fixed edge, the reversal of the magnetic field would lead to current flow in the opposite direction. 

Let us consider now the case where the local spin is also under the action of 
$AC$ 
transverse magnetic field $B_x(t)\equiv2\hbar \Omega/(g\mu_B)\cos(\omega t)$, 
where $\Omega$ is known as the {\em Rabi frequency} or {\em flop rate}.
When the local spin is driven by $B_x(t)$ 
 with the frequency $\omega$ close enough to the natural frequency $\omega_0$, the non-equilibrium occupations $P_i$ can deviate significantly from their equilibrium counterpart $P_{i}^{\rm eq}$. In particular, for a two level system the occupation imbalance $\Delta P=P_{0}-P_{1}$ is given by the steady state solution of the Bloch equations.~\cite{Abragam_Bleaney_book_1970,Delgado_Rossier_pss_2017} Thus, using the definition of $\delta P$, we can write
\begin{equation}
\delta P=\Delta P^{\rm eq}
\frac{\Omega^2 T_1 T_2}{1+ \delta^2T_2^2+ \Omega^2 T_1 T_2},
\label{Bloch}
\end{equation}
where $\Delta P^{\rm eq} \equiv \tanh\left( \beta\hbar\omega_0/2\right)$ is the equilibrium population imbalance and $\delta=\omega-\omega_0$ is the frequency detuning.
In addition to the equilibrium imbalance, 
the non-equilibrium occupation difference and therefore, the induced electrical current, depends on the Rabi flop rate and the two characteristic time scales, 
the longitudinal relaxation time $T_1=1/(\Gamma_{0\to 1}+\Gamma_{1\to 0})$
and the decoherence time $T_2$, also known as the transversal relaxation time in the language of the macroscopic Bloch equations.~\cite{Abragam_Bleaney_book_1970}
If we make the substitution of Eq. (\ref{Bloch}) into the current expression (\ref{current1}), we get
\begin{equation}
I=I_0\Delta P^{\rm eq}\frac{\Omega^2 T_1 T_2}{1+\delta^2 T_2^2+\Omega^2 T_1 T_2},
\label{main0}
\end{equation}
where 
\begin{equation}
I_0=\frac{e}{2T_1}.
\label{I0}
\end{equation}
Equation (\ref{main0}) is the main result of this paper.  It predicts a $DC$ current flowing at the edge of a Quantum Spin Hall when a single localized spin is driven with an $AC$ field.

\section{Estimate of maximal $DC$ current \label{secIII}}
The maximal induced $DC$ current is obtained at resonance ($\delta=0$), when the driving frequency matches the Zeeman frequency, and it is given by
\begin{equation}
I_{\rm max}=I_0\Delta P^{\rm eq},
\end{equation}
obtained when $\Omega^2 T_1 T_2\gg 1$ and assuming $T_1$ is entirely due to the Kondo exchange mechanism envisioned in Fig. 1.  The maximal equilibrium spin polarization $\Delta P^{\rm eq}=1$ is achieved only when the low energy spin state is fully occupied, i.e., $\beta\hbar\omega_0\gg 1$, where $I_{\rm max}=I_0$.  In other words, the magnitude of the maximal current is determined by $T_1$ provided $\Omega^2 T_1 T_2\gg 1$. 
In this limit,
 the spin relaxation rate due to Kondo exchange for a single $S=1/2$ spin interacting with the spin-locked edge of a QSHI is given by:~\cite{Delgado_Rossier_njp_2019}
\begin{equation}
\frac{1}{T_1}\approx\frac{(\rho J)^2\pi}{16}\omega_0,
\label{rate}
\end{equation}
where $\rho$ is the density of states at the Fermi energy of the edge electrons. 
 Equation (\ref{rate}) is derived taking $\rho J$ as a small parameter.  Therefore, 
 an upper bound for the $DC$ current is given by
\begin{equation}
I^{\rm theo}_{\rm max}<\frac{e\pi}{32}\omega_0.
\label{ImaxB}
\end{equation}
For a $DC$ field of 1 Tesla ($\omega_0\approx 1.8\times 10^{11}\; s^{-1}$), standard for ESR experiments,  $I_{\rm max}$ is in the nano-Ampere regime for $T\ll 1.3$ K, well within the instrumental state of the art.  We note that nuclear Zeeman splittings  is 3 orders of magnitude smaller than its electronic counterpart, and the hyperfine interaction is at least 3 orders of magnitude smaller than Kondo exchange. Therefore, nuclear spin relaxation rates, that scale with  the square of the hyperfine interaction, will be many orders of magnitude smaller than their electronic counterparts.

Although Eq. (\ref{I0}) naively implies that a $T_1$ as short as possible is desired, 
the inequality 
$\Omega^2 T_1 T_2\gg 1$ must also hold.
 Given that $T_2<2 T_1$, a short $T_1$ requires a large  Rabi coupling $\Omega$.   Thus,  $T_1$ must remain above $1/\Omega$ so  the maximal current criteria is satisfied.  In practice, this leads to the stricter condition:
\begin{equation}
I_{\rm max}<e\Omega.
\label{strickC}
\end{equation}

In conventional ESR experiments, the spin is driven by the $AC$ magnetic field of a microwave. Typically, cavities are used to increase the magnitude of the $AC$ field.  State of the art values for the $AC$ magnetic field in ESR expriments can be larger than 250 mG.\cite{brustolon09}  For a spin $S=1/2$ with $g=2$ this gives $\Omega\simeq 0.7 MHz $ and, from Eq. (\ref{strickC}), $I<120$ fA, well above state-of-the-art current detectors than can detect changes as small as 10 fA.~\cite{Baumann_Paul_science_2015,Yang_Paul_prl_2019}.

%
Larger values of $\Omega$ have been achieved using ESR-STM, where  several different  driving mechanisms other than Zeeman interaction with the $AC$ field  have been proposed.~\cite{Lado_Ferron_prb_2017,Galvez_Wolf_prb_2019,Ferron_Rodriguez_prb_2019,delgado21}
For Ti-H on MgO,  and $S=1/2$ spin system, $AC$ magnetic fields up to 1 mT have been reported,~\cite{Yang_Paul_prl_2019} with an induced Rabi frequency $\Omega/2\pi\sim 10$ MHz in continuous mode, while Rabi frequencies up to 30 MHz have been demonstrated in pulsed ESR-STM~\cite{Yang_Paul_science_2019} or using double resonance under large $AC$ voltages.~\cite{Phark_Chen_arxiv_2021} Moreover, these conditions can be achieved while keeping the $\Omega^2 T_1 T_2$ factor larger than one.~\cite{Willke_Singha_nanolett_2019,Willke_Paul_sciadv_2018,Phark_Chen_arxiv_2021} These rates translates into maximal currents up to $\sim$3 pA,

\section{Limiting factors \label{secIV}}
Condition (\ref{strickC}) is an upper bound for the pumped current generated by the single-spin resonance. In addition to the conditions leading to this maximum current ($\beta\hbar\omega_0\gg 1$ and $\Omega^2 T_1 T_2\gg 1$), there are a few factors that could reduce the efficiency of this resonant pumping. 
For instance, any mechanism that leads to the local spin relaxation without creating of a 2$k_F$ electron-hole pair will decrease the $DC$ current, for a fixed value of the Rabi coupling.  There are several mechanisms that can relax the spin. First, suppose the material-dependent momentum-spin locking axis $z$ is not perfectly aligned with local spin quantization axis $z'$. In that case, exchange interactions will relax the local spin in the forward-scattering channel that entails no current.  For instance,  let us consider a local spin governed by the Hamiltonian $H=D S_z^2 + E(S_x^2-S_y^2)$, integer spin, and $D<0$.  It can be seen~\cite{Delgado_Loth_epl_2015} that transitions between the ground state doublet are generated by the $S_z$ 
 operator. Therefore, the Kondo exchange with the QSHI edge states is via the $S_z \sigma_z(0)$ operator, which can only produce forward scattering spin-conserving transitions.  In general, the quantization direction of the edge state will depend on momentum and it can point in directions different than the normal.\cite{gosalbez11,Peng_Yuan_natcomm_2017}


%

Second, spin-phonon coupling can represent an important source of spin relaxation in paramagnetic crystals,~\cite{Mattuck_Strandberg_pr_1960} including both one-phonon direct relaxation processes, with a typical relaxation rate proportional to $T$ when $\hbar\omega_0\gg k_BT$, and two-phonon Raman and Orbach processes.~\cite{Abragam_Bleaney_book_1970}
%
 Third, spin relaxation of the current-carrying electron-hole, induced by nuclear spins,\cite{Lunde13} by other magnetic impurities, and with other thermally excited electron-hole pairs in bulk states would reduce the resulting current.   Whereas hyperfine interactions are typically weak,  the case where more than one magnetic center is present at the edge deserves future attention.  One one hand,  having $N$ resonating spins enhances the spin pumping. On the other, electron-hole pairs generated by a given spin can be reabsorbed by the others.   
%

%
Another limiting factor would be the formation of a Kondo singlet, that would quench the magnetic moment of the local spin, reducing its effective coupling to the external driving force.

\section{Discussion and Conclusions \label{secV}}

Here we have proposed a mechanism that permits one to envision an electrically-detected single spin resonance of a magnetic impurity coupled to the edge state of a QSHI. We have demonstrated that the spin-momentum locking at the edge states leads to a spontaneous net current when an electron-hole pair is created by the isotropic exchange coupling with a local magnetic moment. If an external $AC$ driving is capable of inducing a departure of the stationary occupations from their equilibrium counterpart, this in turns generates a measurable $DC$ current. 
We have shown that, with state of the art instrumentation, the  upper limit for the generated $DC$ current is given by the Rabi coupling $\Omega$ of the local spin to the $AC$ driving fields and presented a thorough discussion of the limiting factors that could reduce this maximum induced current. 
We estimate that state-of-the-art ESR instrumentation can provide values of 
$\Omega$ that will induce currents within the current sensitivity, with $DC$ currents  well above the few tens of fA.
Finally, we have proposed several physical realizations, such as magnetic adatoms or molecules attached at the border of a QSHI and probed by a ESR-STM.

%
%
%
%
%
%

\section{Acknowledgments}
We acknowledge fruitful discussions with David Soriano
J.F.R.  acknowledges financial support from 
 FCT (Grant No. PTDC/FIS-MAC/2045/2021),
 SNF Sinergia (Grant Pimag),
FEDER /Junta de Andaluc\'ia --- Consejer\'ia de Transformaci\'on Econ\'omica, Industria, Conocimiento y Universidades,
(Grant No. P18-FR-4834), 
and Generalitat Valenciana funding Prometeo2021/017.
and MFA/2022/045
FD and JFR acknowledge funding from
MICIIN-Spain (Grant No. PID2019-109539GB-C41).
%
%
 This work has
been financially supported in part by FEDER funds.
F.D. thanks the hospitality of the Departamento de F\'isica Aplicada at
the Universidad de Alicante.


%

\end{document}